\documentclass[aps, prd, reprint, showpacs, nofootinbib, longbibliography, floatfix,superscriptaddress]{revtex4-1}

\usepackage{amsmath}
\usepackage{amssymb}
\usepackage[draft]{changes}
\usepackage{comment}
\usepackage{enumitem}
\usepackage{graphicx}
\usepackage{dcolumn}
\usepackage{bm}
\usepackage[hyperindex]{hyperref}
\usepackage[utf8]{inputenc}
\usepackage[caption=false]{subfig}

\usepackage[mathlines]{lineno}

\definechangesauthor[color=blue]{SV}

\newcommand{\dd}{\mathrm{d}}

\newcommand{\myfrac}[2]{%
    \setbox0\hbox{$#1$}        
    \dimen0=\wd0               
    \setbox1\hbox{$#2$}        
    \dimen1=\wd1               
    \ifdim\wd0<\wd1            
        \dfrac{#1\hfill}{#2}   
    \else
        \dfrac{#1}{#2\hfill}   
    \fi
}

\newcommand{\ks}{k_{s}}

\newcommand{\hs}{h_{k_{s}}}
\renewcommand{\r}{R_{H_0}}
\newcommand{\h}{H_0}

\newcommand{\EV}[1]{\vert\HFi\vert}

\newcommand{\HFi}{\bm{k}}

\DeclareFontFamily{U}{mathx}{\hyphenchar\font45}
\DeclareFontShape{U}{mathx}{m}{n}{<-> mathx10}{}
\DeclareSymbolFont{mathx}{U}{mathx}{m}{n}
\DeclareMathAccent{\widebar}{0}{mathx}{"73}

\newcommand{\p}{\partial}

\begin{document}
\title{Vector perturbations in bouncing cosmology}

\author{Nelson~Pinto-Neto}\email{nelson.pinto@pq.cnpq.br}
\affiliation{CBPF - Centro Brasileiro de Pesquisas F\'{\i}sicas, Xavier Sigaud st. 150, zip 22290-180, Rio de Janeiro, Brazil}
\affiliation{PPGCosmo, Universidade Federal do Espírito Santo. Av. Fernando Ferrari 514, zip 29075-910, Vit\'oria-ES, Brazil}

\author{J\'ulio C. Fabris}\email{julio.fabris@cosmo-ufes.org}
\affiliation{Núcleo de Astrofísica e Cosmologia (Cosmo-ufes), PPGCosmo \& Departamento de Física, Universidade Federal do Espírito Santo. Av. Fernando Ferrari 514, zip 29075-910, Vit\'oria-ES, Brazil}
\affiliation{National Research Nuclear University MEPhI, Kashirskoe sh. 31, Moscow 115409, Russia}

\author{J\'unior D. Toniato}\email{junior.toniato@ufes.br}
\affiliation{Núcleo de Astrofísica e Cosmologia (Cosmo-ufes), PPGCosmo \& Departamento de Física, Universidade Federal do Espírito Santo. Av. Fernando Ferrari 514, zip 29075-910, Vit\'oria-ES, Brazil}

\author{G.~S.~Vicente}\email{gustavo@fat.uerj.br}
\affiliation{FAT - Faculdade de Tecnologia, UERJ - Universidade do Estado do Rio de Janeiro (Campus Resende), Rod. Pres. Dutra, km 298,
	zip 27537-000, Rio de Janeiro, Brazil}

\author{Sandro~D.~P.~Vitenti}
\email{vitenti@uel.br}
\affiliation{Departamento de F\'{i}sica, Universidade Estadual
	de Londrina, Rod. Celso Garcia Cid, Km 380, 86057-970,
	Londrina, Paran\'{a}, Brazil}

\date{\today}

\begin{abstract}
An old question surrounding bouncing models concerns their stability under
vector perturbations. Considering perfect fluids or scalar fields, vector
perturbations evolve kinematically as $a^{-2}$, where $a$ is the scale factor.
Consequently, a definite answer concerning the bounce stability depends on an
arbitrary constant, therefore, there is no definitive answer. In this paper, we consider a more general situation where the primeval material medium is a non-ideal fluid, and
its shear viscosity is capable of producing torque oscillations, which can
create and dynamically sustain vector perturbations along cosmic evolution. In
this framework, one can set that vector perturbations have a quantum mechanical
origin, coming from quantum vacuum fluctuations in the far past of the bouncing
model, as it is done with scalar and tensor perturbations. Under this
prescription, one can calculate their evolution during the whole history of the
bouncing model, and precisely infer the conditions under which they remain
linear before the expanding phase. It is shown that such linearity conditions
impose constraints on the free parameters of bouncing models, which are mild,
although not trivial, allowing a large class of possibilities. Such conditions
impose that vector perturbations are also not observationally relevant in the
expanding phase. The conclusion is that bouncing models are generally stable
under vector perturbations. As they are also stable under scalar and tensor
perturbations, we conclude that bouncing models are generally stable under
perturbations originated from quantum vacuum perturbations in the far past of
their contracting phase.
\end{abstract}


\maketitle
\begin{sloppypar}

\tableofcontents

\section{Introduction}

In standard hot big bang cosmology, classical primordial perturbations around a
homogeneous background would never have been in causal contact and structure
formation cannot be explained. Cosmic inflation solves this problem by
generating primordial perturbations of quantum-mechanical origin, which are
later stretched by expansion and explain the observed spectrum of
perturbations~\cite{Planck}. However, in addition to the quantum production of
perturbation modes from vacuum fluctuations, cosmic inflation is preceded by an
initial singularity, at which quantum effects are expected to be relevant.
Therefore, it is natural to ask for a quantum description for both background
and perturbations.

A quantum treatment of the early Universe enables the avoidance of the initial
singularity. The absence of singularities allows the connection of the present
expanding phase to a preceding contracting phase through a bounce \cite{Tipler,
	Brandenberger:1988aj, Lemos:1995qu, AcaciodeBarros:1997gy, Colistete:2000ix,
	Alvarenga:2001nm, Khoury:2001wf, Donagi:2001fs, Finelli:2001sr, Fabris:2002pm,
	Ashtekar:2006rx, agullo2016loop}. The bounce physics depends on the quantization
scheme. In the context of the Wheeler-DeWitt quantization of minisuperspace
models using the de Bohm-de Broglie quantum theory~\cite{Bohm:1951xw,
	holland_1993, Durr:2010}, the Bohmian evolution of the scale factor is free of
singularities: they describe universes that contract classically from infinity,
perform a quantum bounce, and are subsequently ejected into an expanding phase,
where classical evolution, compatible with observations, is rapidly recovered
\cite{AcaciodeBarros:1997gy, Colistete:2000ix, Alvarenga:2001nm,
	Pinto-Neto:2013toa}.

The quantum theory of linear cosmological perturbations can be extended to such
backgrounds~\cite{Pinho:2006ym,Peter:2006hx, Vitenti2013, Falciano2013,
	Peter:2015zaa,Bacalhau:2017hja}. Primordial perturbations can naturally arise
from quantum vacuum fluctuations in the far past of the contracting phase, where
space-time is almost flat, and an adiabatic vacuum state can be prescribed.
These perturbations are amplified during cosmic evolution, becoming the seeds of
the large-scale structures of the Universe in the expanding phase. As well as in
cosmic inflation, scalar and tensor perturbations of quantum mechanical origin
can be shown to be almost scale invariant if the contracting phase is dominated
by a dust fluid (maybe dark matter) at large scales. Furthermore, it can be
shown that they never leave the linear regime up to the expanding phase, where
they necessarily must become non-linear in order to develop structures in the
Universe~\cite{Sandro:tese, Vitenti2012, Pinto-Neto2014}.

In the references cited above, the matter content of the models are described by
perfect fluids or scalar fields. In this case, vector perturbations, evolve like
$a^{-2}$, as usual, where $a$ is the scale factor. For big bang models with
inflation, it is expected that such primordial vector perturbations become
completely negligible after the inflationary phase. However, bouncing models
contain a contracting phase, where these perturbations can increase, and one may
wonder whether they can become non-linear and destroy the homogeneity
or the isotropy of the background while the Universe reaches the
bounce. If one keeps restricted to perfect fluids and/or scalar
fields, the vector modes do not have scale dependent dynamics, and consequently, the answer to this question will depend on an arbitrary constant, hence all answers are possible. However, if we enlarge the possibilities and
consider the primeval material medium as a non-ideal fluid, the shear viscosity
is capable of producing torque oscillations, which can create and sustain vector
perturbations along cosmic evolution. Furthermore, as for scalar and tensor
perturbations, one can assume that vector perturbations also have a quantum
mechanical origin, as described in Ref.~\cite{Grishchuk:1993ab}.

The aim of this paper is to apply the framework developed in
Ref.~\cite{Grishchuk:1993ab} for vector perturbations to the quantum bouncing
models described above. A natural initial adiabatic quantum vacuum state for the
vector perturbations can now be prescribed, which turns possible to evaluate the
evolution of vector perturbations without any arbitrariness. Demanding that they
stay linear during cosmic evaluation imposes constraints on the free parameters
of the background model. We will see that these constraints are
mild, although not trivial. Another important outcome is to evaluate whether
such vector perturbations can provide some signature of the collapsing phase,
seed large-scale cosmic magnetic
fields~\cite{Matarrese:2004kq,Kobayashi:2007wd}, and polarization of the Cosmic
Microwave Background (CMB) spectrum~\cite{Mollerach:2003nq}.

The paper is organized as follows. In Sec.~\ref{sec:non-ideal-fluid}, the
hydrodynamics of non-ideal fluids in general relativity is described. In
Sec.~\ref{sec:pert}, we set up the theory of cosmological perturbations for
linear vector perturbations, taking into account the effects of shear viscosity,
which is responsible for producing torque oscillations. Section~\ref{sec:bounce}
introduces the quantum bouncing model. The formalism described in
Sec.~\ref{sec:pert} is applied to it and the fundamental equations describing
the evolution of vector perturbations are obtained. The consistency conditions
for linearity are analyzed in Sec.~\ref{sec:consist}. In Sec.~\ref{sec:results},
after imposing adiabatic vacuum initial conditions for the vector perturbations,
the analytical results are obtained, which are then confronted with more
detailed numerical calculations. The constraints on the
background model parameters, coming from the linearity
conditions, are also obtained. Finally, in Sec.~\ref{sec:conclusions}, we draw
some general conclusions about our results.

\section{Non-ideal fluid}\label{sec:non-ideal-fluid}

In cosmology, it is usual to consider only perfect fluids as the matter content
of the Universe. However, a realistic model must take into account dissipative
phenomena, which are always present in the macroscopic description of a system.
Bulk and shear viscosity, besides heat flow, are some examples of such
dissipative processes. Applications of bulk viscosity in cosmological models
have a very large literature. These applications begun, to our knowledge, with
the seminal work by Murphy \cite{Murphy:1973zz}, concerning the primordial
universe, and it has been extended to the study of the dark sector of the
Universe (see Ref.~\cite{Colistete:2007xi} and references therein). In the case
of the primordial Universe, they may lead to the avoidance of the initial
singularity; in the case of the dark sector of the Universe, bulk viscosity
effects may imply negative pressure and contribute to the acceleration of the
universe.

Contrary to bulk viscosity, shear viscosity does not affect isotropic and
homogeneous backgrounds. However, at the perturbative level, it has been shown
that shear viscosity can be as important - or even more - as bulk viscosity.
These surprising results have been shown first in the context of warm inflation
\cite{BasteroGil:2011xd,BasteroGil:2012zr}, and the late Universe
\cite{Barbosa:2017ojt,Barbosa:2018iiq}. For the present universe, dissipative
effects may cure some problems connected with the excess of power in matter
agglomeration at small scales, due to the zero pressure of cold dark matter.

The extra piece of the energy-momentum tensor containing bulk and shear
viscosity, as proposed in Refs.~\cite{Eckart:1940te, landau2004fluid}, reads,
\begin{align}
\label{landau-eckart}
\Delta T_{\mu\nu}  &= 2\lambda\sigma_{\mu\nu}\nonumber + \zeta u^\rho{}_{;\rho}\left(g_{\mu\nu} - u_\mu u_\nu\right), \\
\sigma_{\mu\nu} &\equiv u_{(\mu;\nu)} - u_{(\mu} u^\rho u_{\nu);\rho}-\frac{u^\rho{}_{;\rho}}{3}
\left(g_{\mu\nu} - u_\mu u_\nu\right), 
\end{align}
In this expression, $\lambda$ is the shear viscosity coefficient,
$\zeta$ the bulk viscosity coefficient, $g_{\mu\nu}$ the metric, `$;$'
the covariant derivative compatible with the metric, $u^{\mu}$ is the normal
vector orthogonal to the spatial hypersurfaces and $\sigma_{\mu\nu}$ the shear. Round brackets in the indices indicate symmetrization and we are working with a metric signature $(1,-1,-1,-1)$. The explicit form of $\lambda$ and
	$\zeta$, with their dependence on the physical parameters,
depends on the physical system to be considered. This formulation is non-causal,
in the sense that equilibrium is achieved instantaneously. In an isotropic and
homogeneous cosmological background, these parameters normally depend only on
the energy density. However, this is not the case when heat flux is present.

A causal formalism, taking into account a finite speed of sound, has been implemented by Israel and Stewart \cite{Israel:1979wp}. The general expressions, including bulk and shear viscosity, imply transport equations to compute the viscous pressure. In Ref.~\cite{Colistete:2007xi}, the causal formulation of bulk viscosity has been investigated as a description of the dark sector of the Universe. The more important challenge in using the causal formulation is to have a suitable description of the relaxation time, and non-adiabatic sound speed. Strictly speaking, this implies to have a microscopic model for the fluid content. In doing so, hypothesis must be made on the nature of the fluids composing the dark sector of the primordial fields and matter in the Universe.
Alternatively, a phenomenological hypothesis can be implemented. In Ref. \cite{Colistete:2007xi} it has been shown that, in doing so, the results are quite similar to those obtained using the non-causal formalism. Based on these considerations, in what follows we will restrict the analysis to the Eckart-Landau formulation of dissipative phenomena, set down by Eq.~\eqref{landau-eckart}.

The total dissipative energy-momentum tensor can be written as,
\begin{equation}
\Delta T^{\mu\nu} = T^B_{\mu\nu} + T^S_{\mu\nu}.
\end{equation}
The bulk viscosity term alone is given by,
\begin{equation}
T_{\mu\nu}^B  = \zeta u^\rho_{;\rho}(g_{\mu\nu} - u_\mu u_\nu),
\end{equation}
depending essentially on the volume expansion given by $u^\rho_{;\rho}$. The trace part is given by,
\begin{equation}
T^B = 3\,\zeta u^\rho_{;\rho}.
\end{equation}
The shear viscosity term alone is given by,
\begin{align}\label{Tshear}
T_{\mu\nu}^S  &=2 \lambda\sigma_{\mu\nu}.
\end{align}
This term is zero for an isotropic expansion, as we can expect from a shear process.
Naturally, the trace of the shear energy-momentum tensor is zero:
\begin{equation}
T^S = 0.
\end{equation}

Let us make a final remark concerning the Hamiltonian and Lagrangian
formulations of gravitational systems in the presence of dissipative phenomena.
The construction of the energy-momentum tensor including dissipative process
involves thermodynamical arguments. Since, from the pure mechanical and
macroscopic point of view, a dissipative process implies non conservation of the
mechanical energy, with mechanical energy dissipating through heat, the
construction of a Lagrangian and Hamiltonian for dissipative systems is not
always possible. In some cases, this difficulty can be overcome using the
Rayleigh dissipative function \cite{whittaker1988a}, which can be done only when
the dissipative process depends on the velocity, like in the air resistance
phenomena. Indeed, in this case, it is possible to modify the Lagrange equations
as
\begin{equation}
\frac{d}{dt}\frac{\partial {\cal L}}{\partial \dot q} - \frac{\partial {\cal L}}{\partial q} = F(\dot q).
\end{equation}
In general, however, it is not possible to define the Lagrangian of dissipative
systems just by introducing such dissipative functions. Fortunately, a full
Hamiltonian/Lagrangian formulation of the problem we are investigating in the
present article is not necessary, since we are only interested in the linear
perturbative level, in which a straightforward Hamiltonian can be defined, as it
will be seen in the sequel.

\section{Cosmological vector perturbations}\label{sec:pert}

We want to investigate the behavior of small deviations of a given background
cosmology. The geometry of spacetime is then given by
\begin{equation}\label{metric}
g_{\mu\nu}=g^{(0)}_{\mu\nu}+ h_{\mu\nu},
\end{equation}
where $g^{(0)}_{\mu\nu}$ is assumed to be the Friedmann-Robertson-Walker metric
with a flat spatial section. We will work in the synchronous gauge, where
$h_{0\mu}=0$ and, once our interest is in vector perturbations, we can write
\begin{equation}
h_{ij}= a^{2}(\partial_{j}F_{i}+ \partial_{i}F_{j}),
\end{equation}
with $a$ being the background scale factor and $F_{i}$ is an arbitrary vector field satisfying $\p_{i}F^{i}=0$. Latin indices run from $1$ to $3$, indicating spatial components. The perturbed line element is then written as,
\begin{equation}\label{line}
ds^2 = a(\eta)^2\left[\dd\eta^2 -  \left(\delta_{ij}- \partial_{j}F_{i} - \partial_{i}F_{j}\right)\dd x^{i}\dd x^{j} \right],
\end{equation}
where $\eta$ is the conformal time. Substituting this into Einstein's tensor and keeping first order terms only, one obtains,
\begin{align}
G^{0}{}_{0} =& \ \frac{3 a'^2}{a^4}\\[1ex]
G^{0}{}_{i} =& - \frac{\nabla^{2}{} F'_{i}}{2 a^2}  \\[1ex]
G^{i}{}_{j} =&- \left(\frac{3 a'^2}{a^4} - \frac{2 a''}{a^3}\right)\delta^{i}_{\,j} \ + \notag \\
& + \left(\frac{\partial^{i}F''_{j}}{2 a^2} + \frac{\partial_{j}{F^{i}}''}{2 a^2} + \frac{a' \partial^{i}F'_{j}}{a^3}  + \frac{a' \partial_{j}{F^{i}}'}{a^3} \right),\label{feqij}
\end{align}
where the symbol $'$ indicate derivatives with respect to the conformal time $\eta$,
and $\nabla^2 = \partial_i\partial^i$ the spatial conformal
	Laplacian.

The total energy-momentum tensor is written as,
\begin{equation}
T_{\mu\nu}=(\rho+p)u_{\mu}u_{\nu}- pg_{\mu\nu}+ T_{\mu\nu}^{S},
\end{equation}
where $\rho$ is the fluid energy density, $p$ its pressure, $u^{\mu}$ is the
four velocity and $T^{\mu\nu}_{S}$ is the shear viscosity component of the
fluid, given in Eq.~\eqref{Tshear}. At the background level the four velocity is
given by $u^{\mu}=(1/a)\delta_{0}{}^{\mu}$, while its perturbation is described by
a vector function, i.e., $\delta u^{\mu}=(1/a)v^{j}\delta_{j}{}^{\mu}$.
Considering only vector perturbations, the components of the total
energy-momentum tensor read,
\begin{align}
T^{0}{}_{0} &= \ \rho,\\
T^{0}{}_{i} &= - \left(\rho + p \right)v_{i},\\
T^{i}{}_{j} &= - p\,\delta^{i}{}_{j} + \frac{\lambda}{a} \left(\partial^{i}v_{j}  + \partial_{j}v^{i} - \partial^{i}F'_{j} - \partial_{j}{F^{i}}'\right).
\label{Teqij}
\end{align}

With the expressions \eqref{feqij}-\eqref{Teqij} of the Einstein and
energy-momentum tensors, the field equations can be directly obtained. The
background dynamics will be given by the usual Friedmann equations. The linear
perturbations $F_i$ and $v^i$ can be decomposed in terms of eigenfunctions of
the three dimensional Laplace's operator, $Q_{i}$, satisfying the equations,
\begin{equation}\label{base}
	\nabla^{2}Q_i=-k^{2}Q_i\,,\quad \p_{i}Q^{i}=0\,.
\end{equation}
We then write,
\begin{equation}\label{modes}
F_{i}=F(\eta)Q_i,\quad \mbox{and} \quad v_i=v(\eta)Q_i.
\end{equation}
Note that, there are two linearly independent vector fields
satisfying Eq.~\eqref{base}. In practice this means that once quantized, we
would have the equivalent to two uncoupled scalar quantum fields. Nonetheless,
since we consider isotropic vacuum states, both modes have equally defined
vacuum states. In practice, we account for these two modes by multiplying the
vector power spectrum by a factor of 2.

Simplifying expressions with the following definitions,
\begin{align}
h(\eta)=&-kF(\eta),\label{h}\\
\omega(\eta)=& -a(\eta)^{2}(\rho+p)v(\eta),\\
\chi(\eta)=& \ \lambda(\eta)a(\eta)k^{2}\left[F'(\eta)-v(\eta)\right],
\end{align}
Einstein's equations can be recast as
\begin{align}
&-\dfrac{kh'}{2}=\kappa\omega,\label{einstein1}\\
&\omega'+\dfrac{2a'}{a}\,\omega+\chi=0\,,\label{einstein2}
\end{align}
where $\kappa=8\pi l_p^2$ is the gravitational coupling constant, with $l_{p}$
being the Planck length. Equations (\ref{einstein1}) and (\ref{einstein2}) lead
to
\begin{equation}\label{cons}
h''+2\dfrac{a'}{a}h'=\dfrac{2\kappa}{k}\chi\,.
\end{equation}
Note that, without shear viscosity, $\lambda(\eta)=0$, which implies that
$\chi=0$, we get $h'\propto 1/a^2$, as usual.

From its definition, $\omega(\eta)$ can be understood as angular momentum.
Following Ref.~\cite{Grishchuk:1993ab}, in the limit of flat space-time,
equation \eqref{cons} represents Newton's second law in its angular version:
torque is the rate of change of angular momentum. Hence, the $\chi$ function can
be interpreted as torque force in the viscous fluid. As usual, we can take it to
be proportional to the angular displacement of a given element of the fluid,
\begin{align}\label{torque}
\chi(\eta)=k^{2}b^{2}\theta(\eta)
\end{align}
where $\omega=\theta'$, and $b^{2}=v_{t}^{2}/c^{2}$, with $v_{t}$ being the
torsional velocity of sound. Note that originally we had 3 free functions,
$\{F,\, v,\, \lambda\}$, or equivalently $\{h,\,\omega,\,\chi\}$, and only 2
dynamical equations. Thus, imposing \eqref{torque} yields an extra condition
which closes the system.

Equations \eqref{einstein1} and \eqref{einstein2} can now be decoupled,
yielding,
\begin{equation}\label{evolh}
h''+\frac{2a'}{a}\,h'+k^{2}b^{2}h=0\,,
\end{equation}
which can also be written as
\begin{equation}\label{mu}
\mu''+\left(k^{2}b^{2}-\frac{a''}{a}\right)\mu=0\,,
\end{equation}
with $\mu=ah=-akF$\,.

Equation \eqref{evolh} (or Eq.~\eqref{mu}) describes the dynamical evolution of
linear vector perturbations due to torque oscillations in the primordial fluid.
These equations have the same form as the dynamical equations for tensor
perturbations (primordial gravitational waves). The difference is that here the
constant $b$ can vary between $0$ and $1$, while for gravitational waves $b=1$.


\section{Vector perturbations in bouncing models}\label{sec:bounce}

The Wheeler-DeWitt quantization of mini-superspace models using the de
Broglie-Bohm quantum theory introduces quantum corrections in the Friedmann
equations which are able to remove the classical initial singularity of the
Standard Cosmological Model. For a general review on this subject, see
Ref~\cite{Pinto-Neto:2013toa}. In the present paper, we will consider a
particular bouncing solution which contains a dark matter and radiation fluids,
see Ref.~\cite{PintoNeto:2005gx} for its quantum origin, and
Ref.~\cite{Celani:2016cwm} for its connection the observable universe. The scale
factor is obtained as a Bohmian trajectory, and it reads
\begin{equation}
\label{Yscalefactor}
Y(\eta_s)\equiv\dfrac{a(\eta_s)}{a_0}=\dfrac{\Omega_{m0}}{4}\,\eta_s^{2} + \sqrt{\dfrac{1}{x_b^{2}}+\Omega_{r0}\,\eta_s^{2}},
\end{equation}
where $a_0$ is scale factor today, $\Omega_{m0}$ and $\Omega_{r0}$ are the usual
dimensionless densities of presureless matter and radiation, respectively, and
$x_b=a_0/a_b$, with $a_b$ being the value of the scale factor at the bounce. We
are using the dimensionless conformal time variable, appropriated to numerical
integrations, namely $\eta_s=(a_0/\r)\eta$, where $\r=1/\h$ is the Hubble radius
today. The scale factor in Eq.~\eqref{Yscalefactor} describes a universe
dominated by dust in the far past. As the universe contracts, radiation
eventually dominates over dust and near the bounce quantum effects become
relevant. The quantum bounce happens, and it is followed by another radiation
and dust phases, which fits the Standard Cosmological Model before
nucleosynthesis.

The parameter $x_{b}$ can be constrained by imposing that the curvature scale at
the bounce, $L_{b}$, should be at least a few orders of
magnitude bigger than the Planck length. This is because the quantum gravity
approach we are using, the Wheeler-DeWitt quantization, must be understood as an
approximation of a more involved theory of quantum gravity, which should be
valid only at scales not so close to the Planck length. One has that,
\begin{equation}\label{lb}
\left. L_{b} \equiv \dfrac{1}{\sqrt{R}}\right\vert_{\eta_s=0} = \left.\sqrt{\dfrac{a^{3}}{6a''}}\right\vert_{\eta_{s}=0},
\end{equation}
where $R$ is the Ricci scalar. Using values of $H_{0}=70
\,\text{km\,\,s}^{\scriptscriptstyle -1}\,\text{Mpc}^{\scriptscriptstyle -1}$
and $\Omega_{r0}\approx8\times 10^{\scriptscriptstyle -5}$, one can find the
upper bound $x_{b} < 10^{\scriptscriptstyle 31}$. Moreover, the bounce should
take place at energy scales higher than the beginning of nucleosynthesis, which
implies $x_{b}\gg 10^{\scriptscriptstyle 11}$. Hence, one gets,
\begin{equation}\label{xblimit}
10^{11}\ll x_{b} < 10^{31}.
\end{equation}

Concerning the vector perturbations, taking into account the quantum formalism,
we are interested in the evolution of the normal modes $h_{k}$'s coming from the
expansion in terms of creation and annihilation operators, which satisfy the equation of motion~\eqref{evolh}, i.e.,
\begin{equation}\label{evol3}
h_k''+\frac{2a'}{a}\,h_k'+k^{2}b^{2}h_k=0\,,
\end{equation}
The Hamiltonian yielding this dynamical equation reads
\begin{equation}
\label{hamiltonian}
\mathcal{H} = \frac{\Pi_k^2}{2m} + \frac{m\nu^2 h_k^2}{2} \;,
\end{equation}
where $m\propto a^2$ and $\nu = kb$. The constant of proportionality in the
``mass'' $m$ can be inferred from the kinetic term of vector perturbations
coming from the Einstein-Hilbert action. This is true since the Poisson algebra (and consequently the operator
algebra) is defined by the kinetic term. It is
given by (see Ref.~\cite{Pinho:2006ym}),
\begin{equation}
\label{action-vector}
S = \int {\rm d}^4 x \frac{a^2}{16\pi l_p^2}\frac{h'^2}{2},
\end{equation}
where we are using natural units $\hbar=c=1$ and $\eta$ has dimensions of length. Hence $m=a^2/(16\pi l_p^2)$.

Prescribing adiabatic vacuum initial conditions in the far past of the bouncing model yields, see Ref.~\cite{Sandro:vacuum},
\begin{equation}
\label{adiabatic}
\vert h_k\vert = \frac{1}{\sqrt{2m\nu}} = \frac{4l_p\sqrt{\pi}}{a\sqrt{2kb}},
\end{equation}
with $h_k$ having physical dimensions of ${\rm length}^{3/2}$, as it should be.

We now introduce new dimensionless variables, compatible with the expression \eqref{Yscalefactor}:
\begin{align}\label{dimensionless}
\ks=\frac{k \r}{a_0},\; \vert\hs\vert=\sqrt{\frac{a_0^3}{16\pi l_p^2\r}}\vert h_{k}\vert = \frac{1}{Y\sqrt{2k_s b}},
\end{align}
where $k_s$ is the wave number in Hubble radius units. This
	expression accounts for the adiabatic initial condition and must be evaluated
	where the adiabatic approximation is valid. The dynamical equation for the
dimensionless normal modes preserves the form of Eqs.~\eqref{evolh} and
\eqref{mu},
\begin{equation}\label{evolh2}
\hs''+\frac{2Y'}{Y}\,\hs'+\nu_s^{2}\hs=0\,,
\end{equation}
\begin{equation}\label{muk}
\mu_{k_s}'' +\left(\nu_s^{2}-V\right)\mu_{k_s}=0,
\end{equation}
where $\mu_{k_s} = Y\hs$, we have defined an effective wave vector $\nu_s=\ks b$, $V=Y''/Y$ is the potential, and the upper prime now denotes a derivative with respect to $\eta_{s}$.

\section{Consistency conditions}\label{sec:consist}

In this section we will determine the general conditions under which cosmological vector perturbations remain negligible with respect to the background structure.

We start by considering the perturbations in the metric structure. From the line element \eqref{line}, one has that
\begin{equation}
\vert 2\p_{(i}F_{j)}\vert \ll \vert \delta_{ij}\vert .
\end{equation}
Considering a unitary spatial vector field $U^{\alpha}$, we can construct
appropriated scalar quantities from the condition above. In fact, one can see
that,
\begin{equation}
U^{i}U^{j}\p_{(i}F_{j)}=-\,\dfrac{h}{k}\,U^{i}U^{j}\p_{(i}Q_{j)},
\end{equation}
after invoking the definitions \eqref{base} and \eqref{h}. Note that,
$U^{i}U^{j}\p_{(i}Q_{j)}$ is proportional to $k\cos\phi\cos\psi$, where $\phi$
and $\psi$ are the angles between the vector $U^{i}$ and the vectors $k^{i}$ and
$Q^{i}$, respectively. As we noted before, $F_i$ must be expanded
	in terms of two linearly independent eigenfunctions, $Q_i^{(1)}$ and
	$Q_i^{(2)}$, however, here it would only affect our results by a factor of $2$.
	Since at this stage we are only interested in the order of magnitude, it is safe to
	ignore this detail in the present analysis. Thus, it is immediate to see that $\vert
U^{i}U^{j}\p_{(i}Q_{j)}\vert \lesssim \vert k\vert$. This leads to the scalar
condition,
\begin{equation}\label{hcl}
\vert h\vert\ll 1.
\end{equation}
As the perturbation is quantized, the above condition implies that the mean
value of the operator ${\hat{h}}^2$ should be less than unity, yielding,
\begin{equation}
\langle h^2 \rangle=\dfrac{1}{(2\pi)^{3}}\int_{k_\mathrm{min}}^{k_\mathrm{max}}\vert h_{k}\vert^{2}\dd^{3}k \ll 1,
\end{equation}
where $k_\mathrm{min}$ and $k_\mathrm{max}$ are the ultraviolet and infrared
limits, which we will discuss further. Introducing the dimensionless  variables
defined in Eq.~\ref{dimensionless}, one gets
\begin{equation}\label{cond1}
\langle h^2 \rangle=\dfrac{8}{\pi}\frac{l_p^2}{R_{H_0}^2}\int_{k_{s,\mathrm{min}}}^{k_{s,\mathrm{max}}}\dd k_s k_s^2 \vert h_{k_s}\vert^{2}\ll 1.
\end{equation}

We now investigate the role of vector perturbations in the Einstein's equations. Vector perturbations affect the dynamical Einstein's equation through the time derivative of the extrinsic curvature $K_i{}^j$, which contains a background and a first order part
\begin{equation}\label{dotKb}
\partial_t K_i{}^j = \partial_t H \delta_i{}^j - \frac{H {\delta\sigma}_i{}^j}{3} + \dots,
\end{equation}
where $t$ is cosmic time, $H$ is the Hubble function, and the shear tensor reads
\begin{equation}\label{shear}
\delta\sigma_{ij}=K_{ij}-\dfrac{g^{ab}K_{ab}}{3}\,g_{ij},
\end{equation}
which is null in the background.
Hence, from Eq.~\eqref{dotKb} one gets the second condition (see Ref.~\cite{Sandro:tese}),
\begin{equation}\label{C2}
\vert \delta\sigma_i{}^j\vert \ll \left\vert \dfrac{\partial_t H \delta_i{}^j}{3 H}\right\vert ,
\end{equation}
Using the line element \eqref{line} and the decomposition \eqref{modes}, one sees that,
\begin{equation}\label{shear1}
\delta\sigma_{ij}= -\dfrac{a\partial_{\eta} h}{k}\p_{(i}Q_{j)}.
\end{equation}
Multiplying, as before, condition \eqref{C2} by $U^{i}$ in order to construct a consistency scalar relation, one gets,

\begin{equation}
\label{C1}
\left\vert\dfrac{3H \partial_t h}{\partial_t H}\right\vert\ll 1 \Longrightarrow \left\vert\dfrac{\sqrt{\rho} \partial_{\eta} h}{\sqrt{6\pi}l_p(\rho+p)}\right\vert\ll 1 .
\end{equation}
In order to obtain the second form of condition \eqref{C1}, we have used the classical Friedmann equation. Note, however, that quantum effects are important at the background level near the bounce, hence these two forms of the condition are not always equivalent. As the quantum corrections do not modify the matter equation of state relating the pressure and the energy density, $p=w\rho$, and their functional relation with the scale factor, the second form is the one which is valid at all times, including the bounce.

The quantum version of the classical condition \eqref{C1} reads,
\begin{equation}
\label{Cq}
\langle(\partial_{\eta} h)^2\rangle\ll\frac{6\pi l_p^2(\rho+p)^2}{\rho}.
\end{equation}
Using the dimensionless variables, and the power spectrum of vector perturbations
\begin{equation}\label{psv}
P_{v}(\ks)=\frac{\ks}{2\pi^2}\left\vert \hs' \right\vert^2,
\end{equation}
the final form of the second condition reads

\begin{equation}\label{cond2}
\dfrac{4\pi Y^{2}\,l_{p}^{2}}{\Omega_{r0}\r^2}\int_{k_{s,\mathrm{min}}}^{k_{s,\mathrm{max}}}\dd\ks\ks P_{v}(\ks)\ll 1.
\end{equation}

Equations \eqref{cond1} and \eqref{cond2} are the main results of this section.
They are quite general, valid for many theoretical models beyond the one
considered here.

The cosmological model we are considering in this work has two
additional free parameters: $x_{b}$, related with the size of the
bounce, and $b$, the torsional velocity of the sound. The domain of $x_b$ is
given in Eq.~\eqref{xblimit}, while $0<b<1$. The consistency conditions
\eqref{cond1} and \eqref{cond2} will impose further constraints on these
parameters, which will be obtained in the following section.


\section{Results}\label{sec:results}
In this section we analyze the propagation equation of the quantum vector modes given in \eqref{evolh2} and/or \eqref{muk}, with the scale factor \eqref{Yscalefactor}. We will start by doing analytical calculations, which will be confirmed by the numerical results. From now on, we will omit the sub-index $s$ to the redefined variables discussed in the previous sections.

Working with the current values $\Omega_{r0}\sim 10^{-4}$ and $\Omega_{m0}\sim 0.274$, equation \eqref{Yscalefactor} becomes,
\begin{equation}
Y=6.85\times 10^{-2}\eta^2 + \sqrt{\frac{1}{x_b^2}+\frac{\eta^2}{10^4}}.
\end{equation}
The production of vector perturbations depends on the influence of the potential $V$ over the effective wave number $\nu$ in Eq.~\eqref{muk} of perturbation modes. In different conformal times, the potential will behave according with the dominant phase at that epoch. Far from the bounce, dust is the dominant component, when $\vert\eta\vert\gg 0.15$ and the potential assumes the form,
\begin{equation}\label{dust}
V\approx\frac{2}{\eta^2},\quad \mbox{for} \ \ \vert\eta\vert\gg 0.15.
\end{equation}

For small values of the conformal time, radiation begins to dominate. This phase is divided in two parts. The first one is when the quantum effects are still sub-dominants, resulting,
\begin{equation}
V\approx\frac{13.7}{\eta},\quad \mbox{for} \ \ \frac{9}{x_b^{2/3}}\ll \eta \ll 0.15.
\end{equation}
After that, quantum effects become important and one has,
\begin{equation}
V\approx\frac{10^4}{x_b^2 \eta^4},\quad \mbox{for} \ \ \dfrac{10^{2}}{x_b^{1}}< \vert\eta\vert < \dfrac{9}{x_b^{2/3}}.
\end{equation}
For even smaller values of $\vert\eta\vert$ the quantum effects are completely dominant during the bounce phase, with
\begin{equation}\label{vbounce}
V\approx \frac{x_b^2}{10^4}, \quad \mbox{for} \ \ 0 \leq \vert\eta\vert < \dfrac{10^{2}}{x_b}.
\end{equation}

As mentioned in Section \ref{sec:bounce}, we will prescribe adiabatic vacuum initial conditions, given in Eq.~\eqref{adiabatic}. In terms of $\mu$ they read,
\begin{equation}\label{vacuum}
\mu_{ini}=\dfrac{e^{-i\nu\eta}}{\sqrt{2\nu}},
\end{equation}
which should be imposed at the asymptotic past, far from the bounce, where dust
dominates. Taking into account \eqref{dust} and the initial condition above, the
solution of \eqref{muk} in the dust phase is recasted to be
\begin{equation}\label{mudust}
\mu_k(\eta)=-\frac{\sqrt{\pi \eta}}{2}\,H_{-3/2}(\nu\eta),
\end{equation}
	with $H_{n}$ being the Hankel function of type one. In the region where
$\nu\eta\ll 1$, but still in the dust phase, where
$Y\propto\eta^2$, we can expand formally the solution above in
	powers of $\nu$. The Hankel function is a combination of two
Bessel functions ($J_{3/2}$ and $J_{-3/2}$) and each Bessel function can
be expanded in terms of a power-law times a power series in its argument
squared. Consequently, when expanding \eqref{mudust}, we have two distinct power series, each multiplying a different power-law. In terms of $h_k$, they reads
\begin{align}\label{hankel_series}
h_k=& \ \frac{\mu_k}{Y}\nonumber\\
\approx& \ \nu^{3/2}\left[A_1+\mathcal{O}\left(\nu^2\right)\right]
+\nu^{-3/2}\left[\frac{A_2}{\eta^3}+\mathcal{O}\left(\nu^2\right)\right],
\end{align}
where $A_1$ and $A_2$ are constants.

The Hamiltonian leading to Eq.~\eqref{evolh2},
\begin{equation}\label{hamilt}
{\mathcal H}=\dfrac{\Pi_{k}^{2}}{2Y^{2}}+\dfrac{Y^{2}\nu^{2}\,h_{k}^{2}}{2},
\end{equation}
yields the canonical equations
\begin{equation}\label{canonicaleq}
h_{k}'=\dfrac{\Pi_{k}}{Y^{2}}\,,\quad \Pi_{k}'= -Y^{2}\nu^{2}h_{k}.
\end{equation}
For small values of $\nu$, these equations can be solved in an iterative manner, reproducing the power series discussed above, with the leading order giving,
\begin{align}\nonumber
h_k =& \ C_1(\nu)\left[1+\mathcal{O}\left(\nu^2\right)\right]\\ \label{expansion}
&+C_2(\nu)\left[\int \frac{d\eta}{Y^2}+\mathcal{O}\left(\nu^2\right)\right],\\ \nonumber
\Pi_{k}=& \ C_2(\nu)\left[1+\mathcal{O}\left(\nu^2\right)\right]\\
&+ C_1(\nu)\left[-\int Y^{2}\nu^{2}d\eta+\mathcal{O}\left(\nu^4\right)\right].
\end{align}
Specifying for the dust case ($Y\sim \eta^2$), the matching between \eqref{expansion} and \eqref{hankel_series} gives the $\nu$ dependence of the $C$'s constants above, namely $C_1\propto \nu^{3/2}$ and $C_2\propto\nu^{-3/2}$.

The evolution of these perturbations is exactly the same of tensor perturbations.
Therefore, their power spectrum and spectral index are already known (see, for instance, Ref.~\cite{Pinho:2006ym}) and they satisfy the relation
\begin{equation}
\nu^3\vert h_k\vert^2 \propto \nu^{n_{T}},\quad \mbox{with}\quad n_T=\frac{12w}{1+3w},
\end{equation}
where $w=p/\rho$ is the equation of state parameter of the fluid which is dominating the background when the mode is crossing the potential, $\nu^2 \approx V$. In the case of dust domination, $w=0$, with $Y\propto \eta^{2}$, $\eta_T=0$, one gets the growing modes when the Universe is contracting,
\begin{equation}\label{hdust}
h_k\propto\frac{\nu^{-3/2}}{\eta^3}\,,\quad \Pi_{k}\propto \nu^{-3/2}\,.
\end{equation}
When afterwards radiation dominates the background evolution, $Y\propto\eta$, one gets
\begin{equation}\label{hdustr}
h_k\propto\frac{\nu^{-3/2}}{\eta}\,,\quad \Pi_{k}\propto \nu^{-3/2}\,.
\end{equation}
On the other hand, if the mode crosses the potential already in the radiation domination phase of the contraction, one has $\eta_T=2$ and $Y\propto\eta$, yielding,
\begin{equation}\label{hrad}
h_k(\eta)\propto\frac{\nu^{-1/2}}{\eta},\quad \Pi_{k}\propto \nu^{-1/2}\,.
\end{equation}
During the bounce, the potential is almost constant and nothing happens.

\begin{figure}[t]
	\centering
	\includegraphics[width=\linewidth]{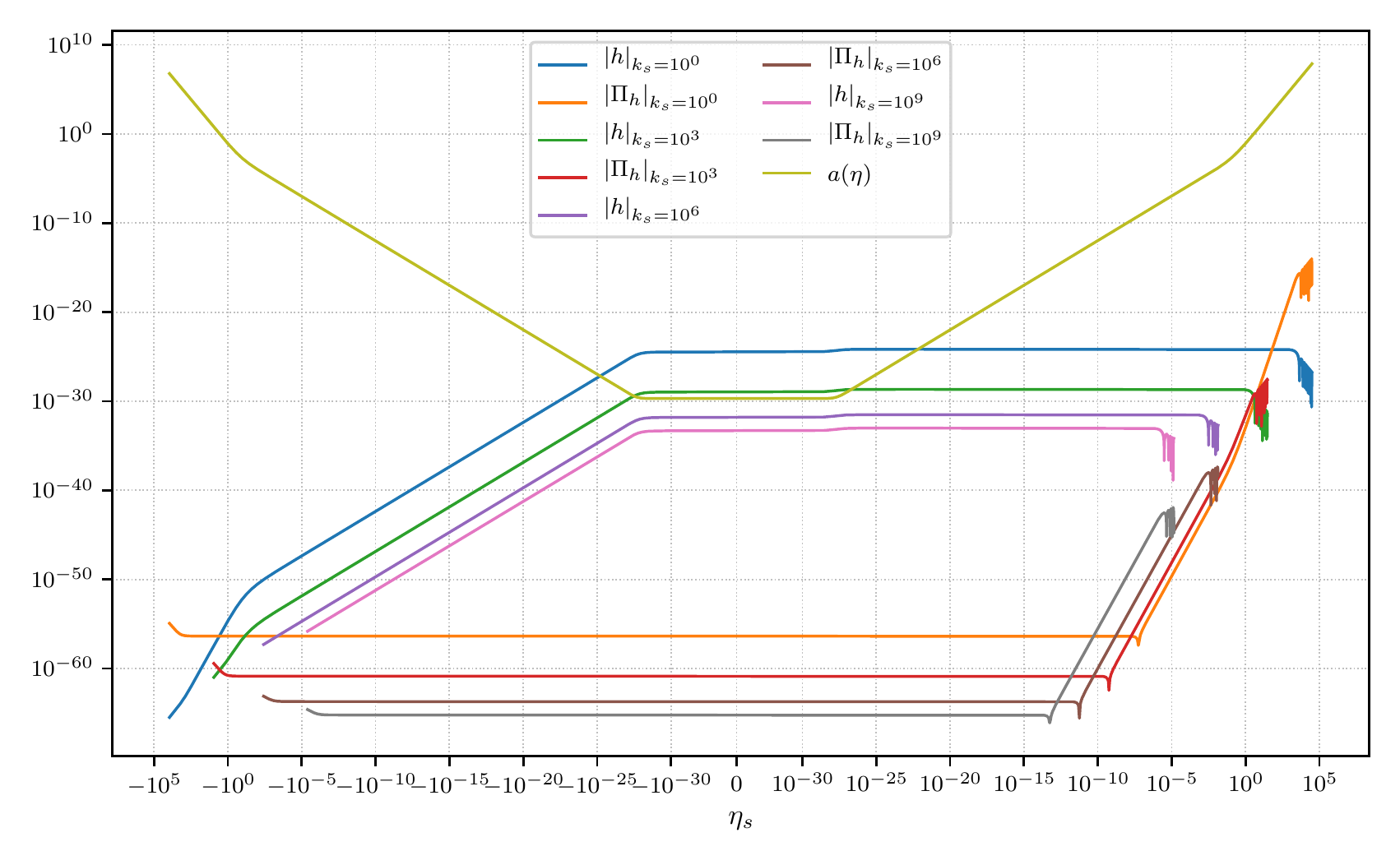}
	\caption{Amplitudes of $h_{k}$ and $\Pi_{k}$ when the crossing of the potential occurs in two different phases. The relation \eqref{hdust} is verified for $k_s$ equal to $10$ and $10^3$, when the modes cross the potential during the dust phase. For the higher values of $k_s$, $10^6$ and $10^9$, the crossing occurs during the radiation phase, satisfying relation \eqref{hrad}. The scale factor is also shown in the figure to a better distinction fo each phase of evolution. The values used for the free parameters are $x_{b}=5\times 10^{29}$ and $b=10^{-3}$.}\label{fig1}
\end{figure}

In the expanding phase, the growing mode of $h_k$ becomes a decaying mode and $h_k$ saturates up to returning to the oscillatory phase, when $\nu^2 > V$. In the case of $\Pi_k$, there is a $\nu^2$ growing mode correction which eventually dominates the constant mode,
and $\Pi_k$ grows up to the oscillatory regime, either with a $k^{1/2}$ spectrum in the case of dust entrance, or $k^{3/2}$ for radiation entrance. The behaviors obtained in \eqref{hdust}, \eqref{hdustr} and \eqref{hrad}, can be verified through the numerical results presented in Fig.~\ref{fig1}, as well as the conclusions relative to the expanding phase.

\begin{figure}[t]
	\includegraphics[width=\linewidth]{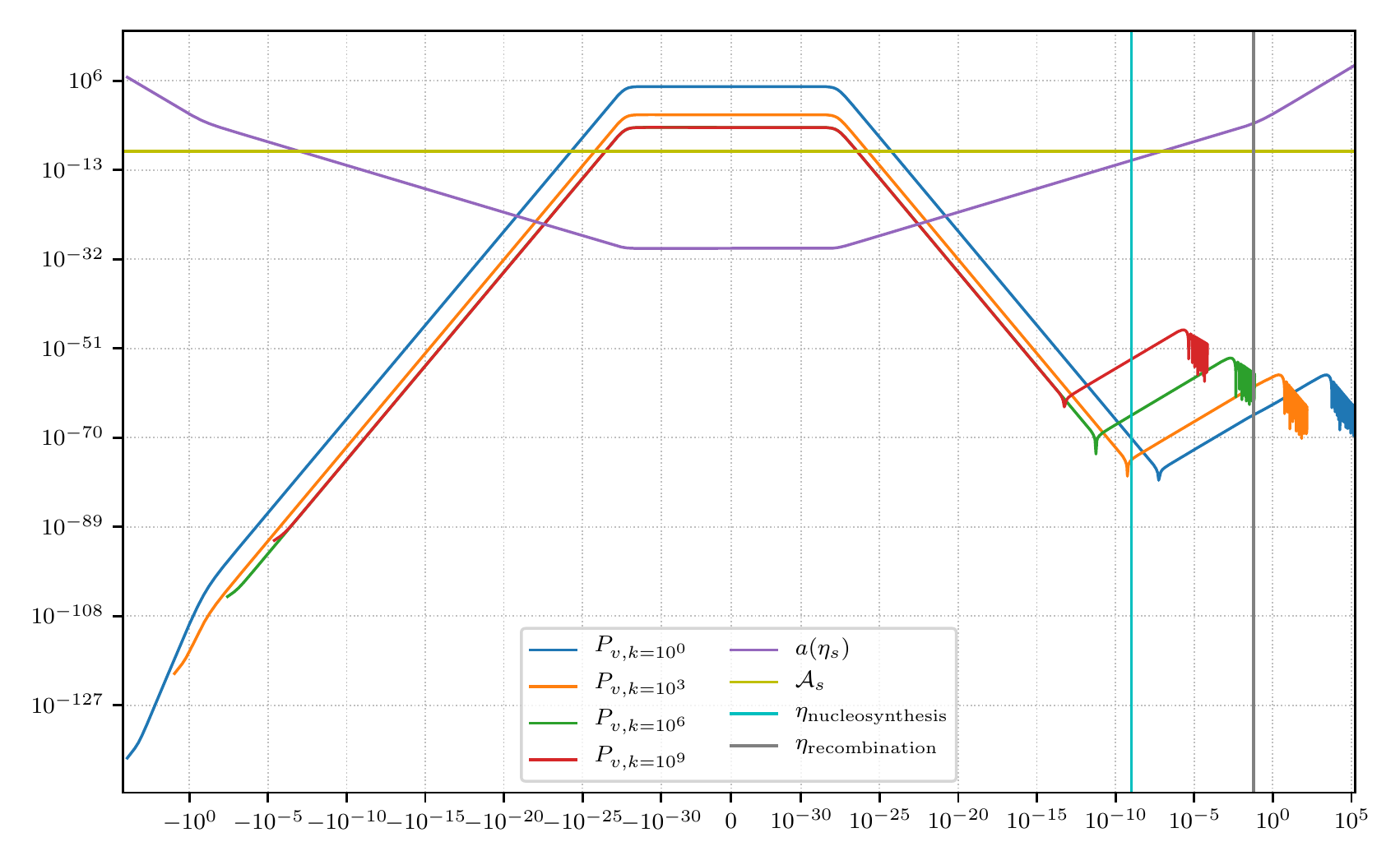} \caption{Behavior of
	the power spectrum $P_{v}$ for distinct values of $k_s$. For the two lowest
	values of $k_s$, $1$ and $10^3$, the vector modes are crossing the potential
	during dust domination phase and relation \eqref{pdust} is observed. The
	highest $k_s$ values, $10^6$ and $10^9$, show the scale independence of $P_v$
	for modes crossing the potential during the radiation phase, in agreement with
	\eqref{prad}. The two vertical lines mark the nucleosynthesis and recombination
	era, while the ${\cal A}_s$ plot show the approximated primordial amplitude for
	scalar modes. The values used for the free parameters are $x_{b}=5\times
	10^{29}$ and $b=10^{-3}$.}\label{fig2}
\end{figure}

The behavior of the power spectrum can be directly inferred considering Eqs.~\eqref{psv} and \eqref{expansion}. When the modes cross the potential in the dust dominated phase, one has,
\begin{equation}
\label{pdust}
P_v\propto\frac{\nu^{-2}}{\eta^8},
\end{equation}
and when they cross in the radiation era, one gets
\begin{equation}\label{prad}
P_v\propto \frac{1}{\eta^4}.
\end{equation}
These behaviors can also be verified numerically in Figure \ref{fig2}.

As it can be seen from Figs.~\ref{fig1} and \ref{fig2}, the most critical region
for conditions \eqref{cond1} and \eqref{cond2} to be satisfied is during the
bounce, where the vector perturbations amplitudes reach their maximum value. Let
us then evaluate these conditions at the bounce. We first need to know how these
quantities are scaled with $x_b$, $b$ and $k$ when $\eta=0$.

The scalings of $b$ and $k$ of both $h_k$ and $P_v$ are embedded in relations
\eqref{hdust}-\eqref{prad}. In the case of $x_b$, note that $h_k$ grows as
$1/\vert\eta\vert$ up to the bouncing phase, which begins in $\eta=-10^2/x_b$
[cf. Eq.~\eqref{vbounce}]. Hence, $\vert h_k\vert \propto x_b$.  For the $P_v$
scaling, note that it can be written as,
\begin{equation}\label{pradPi}
P_v = \frac{\Pi_k^2 k}{Y^4}.
\end{equation}
As $\Pi_k$ is a constant in the contraction, this constant depends only on
$\nu$, and $Y\propto 1/x_b$, hence, $P_v\propto x_b^4$. These results can also
be verified numerically, yielding:
\begin{enumerate}[label=\roman*)]
\item Modes crossing the potential at matter domination phase:
\begin{equation}
h_k\simeq \frac{10^2 x_b }{\nu^{3/2}}\, , \quad P_v\simeq 10^{-124}\frac{x_b^{4}}{b\,\nu^2}.
\end{equation}

\item Modes crossing the potential at radiation domination phase:
\begin{equation}
h_k(\eta)\simeq \frac{x_b }{\nu^{1/2}}\quad , \quad P_v\simeq 10^{-126}\frac{x_b^{4}}{b}.
\end{equation}
\end{enumerate}
Note there is always an extra $b$ in $P_v$ due to its definition, which is proportional to a $k$ factor [cf. Eq.~\eqref{psv}].

We can now perform the integrations in the consistency relations \eqref{cond1}
and \eqref{cond2}. Let us begin with condition \eqref{cond1}. The modes crossing
the potential region during the dust phase have $\nu_c\sim \sqrt{2}/\eta_c$,
where the index $c$ refers to ``crossing". Since the dust domination era ends
when $\eta\sim 0.15$, we than split each integral in two parts, divided by
$\sqrt{2}/0.15\approx 10$. Thus, \eqref{cond1} yields,
\begin{align}
\frac{8}{\pi}\frac{l_p^2}{\r^2 b^3}\int_{\nu_\mathrm{min}}^{\nu_\mathrm{max}} \vert h_k\vert^2\nu^2d\nu= \frac{8}{\pi}\frac{l_p^2 x_b^2}{R_{H_0}^2 b^3}
\bigg[10^2\ln{10} \ -\notag\\
- \ln{\nu_\mathrm{min}} +\frac{(\nu_\mathrm{max}^2-10^2)}{2}
\bigg].
\end{align}
The solution presents an infrared and an ultra-violet divergence. The infrared divergence is logarithmic. As the number in front of the integral is very small, even assuming the minimum value of $b$ ($b > 10^{-26}$, as we will see in the end of this section) leads to an infrared cut-off $L_{\rm infrared}\approx \exp{(10^{24})}\r$, which is beyond any imaginable physical scale.
In the case of the ultra-violet divergence, we use as $\nu_\mathrm{max}$ the value of the maximum of the potential $V$, which happens at the bounce, since modes with $\nu$ beyond this value will only oscillate without being enhanced. Thus, from \eqref{vbounce}, one has $\nu_\mathrm{max}=x_b^2/10^4$. Taking only the dominant term, the consistency condition \eqref{cond1} becomes,
\begin{equation}\label{cond11}
\frac{4}{10^4\pi}\frac{l_p^2 x_b^4}{R_{H_0}^2 b^3}\,\ll\, 1.
\end{equation}
\begin{figure}[t]
\centering
	\includegraphics[width=.9\columnwidth]{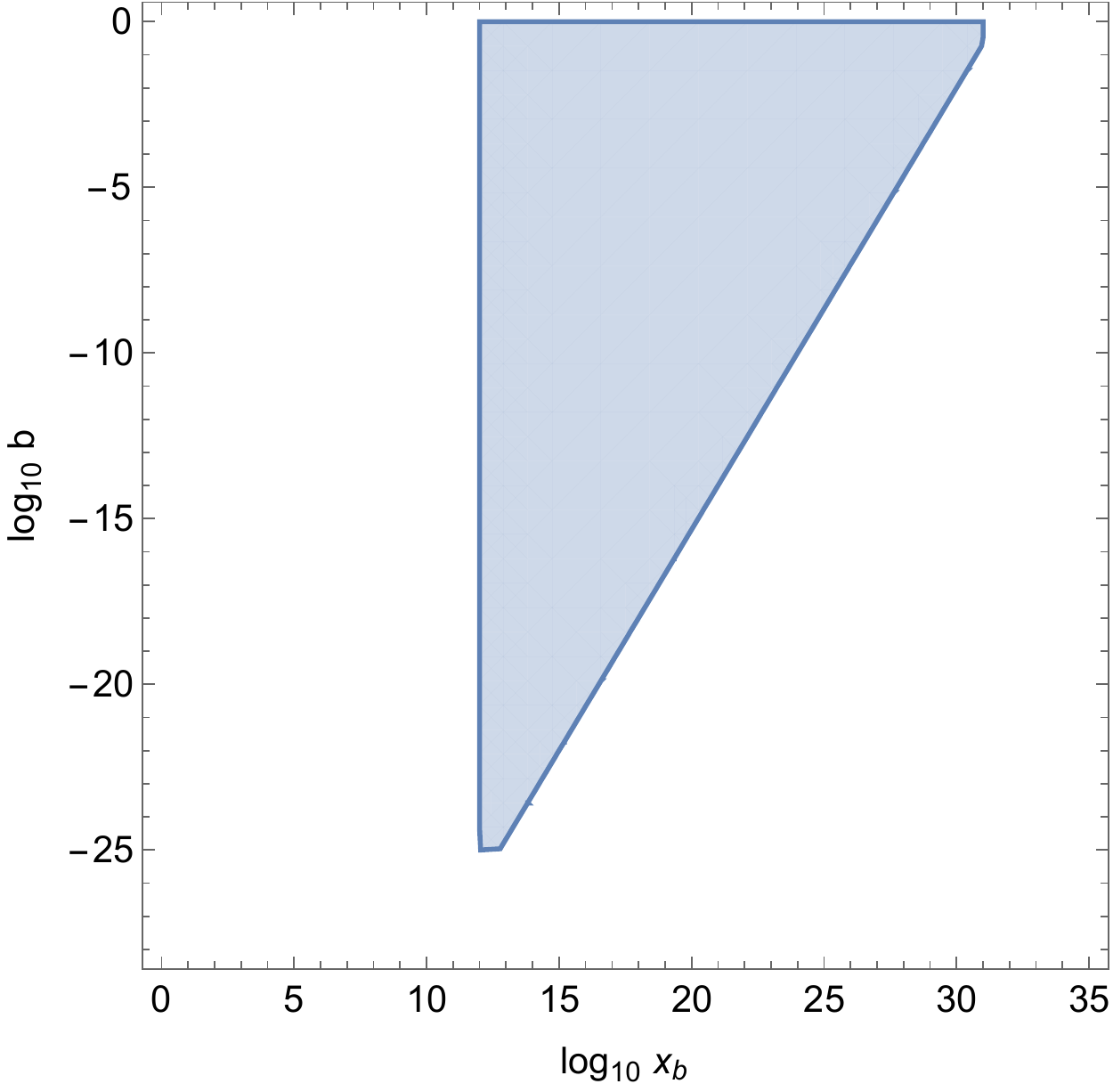}
	\caption{Region of stability in parameter space for the bouncing models considered. The colored area represents the values of $b$ and $x_b$ which satisfies the consistency conditions, following Eq.~\eqref{cond}.}
	\label{fig3}
\end{figure}

For the second condition, given in \eqref{cond2}, the procedure is the same. However it will result in a much less restrictive constraint, as one can infer from the small values of $P_v$ at the bounce, hence it is irrelevant.

Expression \eqref{cond11} reduces to,
\begin{equation}\label{cond}
\frac{x_b^4}{b^3}\,\ll\, 10^{126}.
\end{equation}
Using the limits on $x_b$ given in Eq.~\eqref{xblimit}], and $0<b<1$, the region in parameter space where vector perturbations remain controlled in such bouncing models are shown in Figure \ref{fig3}. Note that the minimum value allowed for $b$ is $b \approx 10^{-26}$.

\section{Conclusions}\label{sec:conclusions}

In this paper we set up the conditions under which linear vector perturbations
remain controlled along the evolution of a general homogeneous and isotropic
cosmological model. We considered a non-ideal fluid, and its shear viscosity is
capable of producing torque oscillations, which can create and dynamically
sustain vector perturbations along cosmic evolution. In this framework, vector
perturbations can be quantized. The resulting conditions \eqref{cond1} and
\eqref{cond2} apply to any cosmological model ruled by Einstein's equations, and
some particular quantum mini-superspace models.

One important application of the established conditions is to investigate
whether bouncing models are stable under vector perturbations. In the case of a
well known quantum bounce, which fit cosmological observations at the background
and linear perturbation level, it was shown that there is a large range of
parameters in which the model is stable. However, as vector perturbations reach
their largest values around the bounce itself, and as they decay afterwards, it
seems to be impossible to detect their fingerprints in cosmological
observations. Hence, vector perturbations, as modeled here,
cannot be used to distinguish this particular bouncing model from inflationary
models.

As bouncing models have already been shown to be stable under linear scalar and
tensor perturbations, the present result indicates that bouncing models are
stable under general linear perturbations as long as initially
the only departure from a homogeneous and isotropic geometry arise from quantum
fluctuations.

\begin{acknowledgments}
	J.C.F. thanks Conselho Nacional de Desenvolvimento Científico e Tecnológico (CNPq) and Fundação de Amparo à Pesquisa e Inovação do Espírito Santo (FAPES) (Brazil) for partial financial support. J.D.T. thanks Coordenação de Aperfeiçoamento de Pessoal de Nível Superior (CAPES) and FAPES (Brazil) for their financial support trough the Profix program. N.P.N. acknowledges support of CNPq (Brazil) under grant PQ-IB 309073/2017-0.
\end{acknowledgments}

\end{sloppypar}

%

\end{document}